\documentclass[12pt]{article}
\usepackage{a4}
\input{psfig}

\begin{document}

\author{Mark Hindmarsh\\[3pt]
Centre for Theoretical Physics\\
University of Sussex\\
Brighton BN1 9QH\\
U.K.\\
e-mail: {\tt m.b.hindmarsh@sussex.ac.uk} }
\date{}

\title{COSMIC STRINGS -- DEAD AGAIN?\thanks{Talk given at 
Cosmo 97, Ambleside, England, Sept 97}}

\maketitle

\begin{abstract}
I report on recent numerical simulations of the simplest field 
theory with cosmic string solutions, the Abelian Higgs model. 
We find that random networks of string 
quickly converge to a scaling solution in
which the network scale length $\xi$ increases linearly with time.
There are very few loops with sizes less than $\xi$, and the 
strings are smooth, showing no signs of ``small scale structure''.
We claim that particle production is 
the dominant energy-loss mechanism, not gravitational radiation as
previously thought.  For strings in Grand Unified Models, 
stringent constraints can be placed from cosmic ray observations 
on the string tension
$\mu$:  we estimate $G\mu < 10^{-9}$, three orders of magnitude lower
than the constraint from Cosmic Microwave Background fluctuations.
\end{abstract}

The reason for studying cosmic strings and other topological defects
\cite{VilShe94,HinKib95} is
that they are possible relics from phase transitions in the hot Big Bang
model, and as such provide one of the few ways of gaining information
about the very early Universe.  The study of cosmic strings has been
built up into a scenario, based on
the notion that a tangled network of strings would have been formed at a
phase transition, and subsequently would have evolved in a
scale-invariant manner to the present day.  
Strings and other defects can leave observable signals in the Cosmic Microwave 
Background fluctuations. Recent work however \cite{DefCMB} tends to discount 
defects as the source of the fluctuations, and limits the string tension $\mu$
in combination with Newton's constant $G$ to $G\mu < 10^{-6}$.

Recent numerical work on cosmic strings \cite{VinAntHin98} 
threatens a radical revision of 
the traditional scenario.  We claim that in the string network loses 
a big (and constant) fraction of its energy into super-massive particles
in every expansion time, which for strings in Grand Unified Theories
(GUTs) would decay into extremely energetic electrons, protons,
$\gamma$-rays and neutrinos.  The observed flux of cosmic rays at above 
$10^{19}$ eV constrains the allowed injection rate of such particles
\cite{BhaRan90,ProSta96,SigLeeCop96},
which is dependent on the mass of the GUT particles, and hence the
string tension is constrained.  
We also claim that there is negligible energy
loss to gravitational radiation, at odds with current belief.

Our simulations use the
lattice formulation of the Abelian Higgs model due to Moriarty, 
Myers and Rebbi \cite{MorMyeReb88}, which is essentially Hamiltonian
lattice gauge theory.  The random initial conditions appropriate to the
high temperature phase of the early Universe are created by drawing the
scalar field $\phi$ from a Gaussian random field distribution.  In the first 
part of the simulation 
then allowed to evolve dissipatively to cool the system and 
eliminate the spurious
high-frequency modes.  

\begin{figure}
\centerline{\psfig{figure=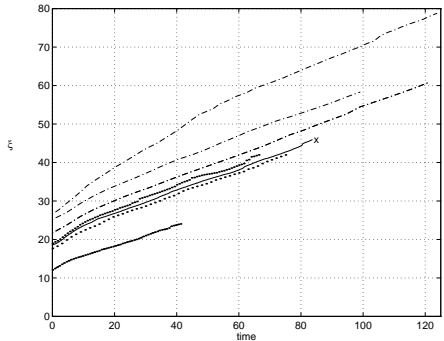,height=1.8in}}
\caption{\label{f:xis.eps}Plots of $\xi_{\rm p}$ for a series of $336^3$
simulations with different lattice spacings $a$.
From top to botttom $a=$0.75, 0.65, 0.75, 0.4,
0.5, 0.45 and 0.25. 
$\xi_{\rm p}$ is given in units of the inverse
scalar mass $m_s^{-1}=$.}
\end{figure}

We are principally interested in how the length of the network of string 
$L(t)$ decays with time: the scaling scenario demands that it be a power
law with exponent -2. An alternative way of phrasing this is to define a
length scale $\xi_{\rm p}(t) = \sqrt{V/L(t)}$, which then should
increase linearly.  (The subscript ``p'' reminds us that we measure the
physical length of string by tracing the zeros, not the invariant
length, which is more usually considered.)  The results are shown in
Figure (\ref{f:xis.eps}).  Note that by ``network'' we mean those
strings whose length is greater than $\xi_{\rm p}$; the rest we count as
loops.  

It is clear that the behaviour of $\xi_{\rm p}$ is extremely linear in
the second half of the simulations. We find 
$
\xi_{\rm p} = x t^p,
$
with $x\simeq 0.3$ and $p=1.00\pm0.03$.  We also find that the strings
are very smooth: there is no sign of any scale 
in the fractal dimension of the string network other than 
$\xi_{\rm p}$.  This is to be contrasted with numerical simulations in
which relativistic strings are simulated directly and there is a
lower cut-off on the allowed loop size \cite{StrSim90}.  
We have argued elsewhere
\cite{VinHinSak97} 
that small-scale structure is a numerical artifact caused by this
cut-off.  The last of our main results is that in our simulations
less than 3\% of the string was in loops.

The real question here is how is the network scaling, and scaling so
accurately?  Somehow, it is losing energy, and it must be losing energy
into radiative modes of the field, as it is not losing it into loops. 
In the traditional string scenario this should not happen.  Perturbative
calculations \cite{PartProd} indicate that the string must be accelerating faster than
the mass of the radiated particle, although these calculations strictly 
apply only when there is no back-reaction on the string as a result 
of the particle emission, i.e.\ when the emitted particle 
is much lighter than the mass scale of the string. This does not turn out to 
give the right scaling law. The mechanism must therefore be non-perturbative.  

The implications of this result for the cosmic string scenario are
profound.  If more than $10^{-3}$ of the energy density of a scaling
network go into GUT mass bosons, all decaying into Standard
Model particles, then the bounds from the observed flux of cosmic rays
of energy above $10^{19}$ eV are violated 
\cite{BhaRan90}.  Using more detailed
calculations \cite{ProSta96,SigLeeCop96}, we derive a limit
\begin{equation}
G\mu < 10^{-9} f_X^{-1.3},
\end{equation}
where $f_X$ is the fraction of the energy ending up as quarks and
leptons.  In the likely case $f_x \sim 1$, this limit is three orders of
magnitude stronger than that from CMB observations.

{\it Note added.}  It has been pointed out \cite{BerBlaVil98} that 
one cannot use UHE cosmic rays to bound strings in this way, as the 
range of gamma rays of this energy is not more than about 20 Mpc, due
to pair production on the microwave background photons.  However, 
there are equally strong bounds from lower energy gamma rays, in the 
range 1-10 GeV, which are produced copiously in a cascade process 
\cite{ProSta96,SigLeeCop96}.  Using these data, one still arrives 
at a bound of about $G\mu < 10^{-9}$ \cite{AntHinVin98}.

MH is supported by PPARC Advanced
Fellowship B/93/AF/1642, and by PPARC 
grants GR/K55967 and GR/L12899.  The work was conducted in cooperation 
with Silicon Graphics/Cray Research using the UK-CCC Origin 
2000, supported by HEFCE and PPARC.




\begin{thebibliography}{99}
\bibitem{VilShe94} A. Vilenkin and E.P.S. Shellard, {``Cosmic Strings
and Other Topological Defects''} (Cambridge Univ. Press, Cambridge, 1994)
\bibitem{HinKib95} M. Hindmarsh and T.W.B. Kibble, {\it Rep. Prog. Phys.}
{\bf 58} 477 (1995).
\bibitem{DefCMB} A. Albrecht, R. Battye and J. Robinson, {\it Phys. Rev. Lett.}
{\bf 79} 4736 (1997);
U.L. Pen, U. Seljak and N. Turok, {\it Phys. Rev. Lett.} {\bf 79} 1611 (1997);
B. Allen et al., {\it Phys. Rev. Lett.} {\bf 79} 2624 (1997).
\bibitem{VinAntHin98} G.R. Vincent, N. Antunes and M. Hindmarsh, 
hep-ph/9708427 (to appear in {\it Phys. Rev. Lett.}).
\bibitem{BhaRan90} P. Bhattacharjee and N.C. Rana, 
{\it Phys. Lett.} {\bf 246B} 365 (1990);
\bibitem{ProSta96} R.J. Protheroe and T. Stanev, 
{\bf 77} 3708 (1996) [ E {\bf 78}, 3420 (1997) ].
\bibitem{SigLeeCop96} G. Sigl, S. Lee and P. Coppi, astro-ph/9604093.
\bibitem{StrSim90} D. Bennett, in 
``The Formation and Evolution of Cosmic Strings'' eds. 
G. Gibbons, S.W. Hawking, and T. Vachaspati, 
(Cambridge Univ. Press, Cambridge, 1990); F. Bouchet, {\it ibid.}; 
E.P.S. Shellard, {\it ibid.}
\bibitem{PartProd} T. Vachaspati, A.E. Everett and A. Vilenkin,
{\it Phys. Rev.} {\bf D30} 2046 (1984);
M. Srednicki and S. Theisen, {\it Phys. Lett.}
{\bf 189B} 397 (1987).
\bibitem{MorMyeReb88} K.J.M. Moriarty, E. Myers and C. Rebbi,
{\it Phys. Lett.} {\bf B207} 411 (1988).
\bibitem{VinHinSak97} G.R. Vincent, M. Hindmarsh and M. Sakellariadou,
{\it Phys. Rev.} {\bf D56} 637 (1997).
\bibitem{BerBlaVil98}  V. Berezinsky, P. Blasi, A. Vilenkin,
astro-ph/9803271.
\bibitem{AntHinVin98} N. Antunes, M. Hindmarsh and G.R. Vincent,
in preparation.
\end{thebibliography}
\end{document}